%Paper: cond-mat/9306031
%From: physics!cdgupta@vigyan.ernet.in
%Date: Thu, 10 Jun 93 19:58:46 EDT

\magnification1200
\footline={\hfil\number\pageno\hfil}
% FIGURES
\newcount\static \static=1
\newcount\FVF \FVF=2
\newcount\ddd \ddd=3
\newcount\ddc \ddc=4

%REFERENCES
\newcount\JA
\newcount\AN
\newcount\VF
\newcount\KW
\newcount\NSE
\newcount\MC
\newcount\DM
\newcount\GO
\newcount\GS
\newcount\KR
\newcount\DAS
\newcount\LE
\newcount\BGS
\newcount\KM
\newcount\WA
\newcount\UY
\newcount\BRH
\newcount\SIG
\newcount\RY
\newcount\PY
\newcount\VM
\newcount\VMO
\newcount\CD
\newcount\FV
\newcount\BY
\newcount\BYP
\newcount\AAY
%EQUATIONS
\newcount\sectni  %section of introduction (I)
\newcount\sectnf  %section with formulae (II)
\newcount\sectnr  %section with methods (III)
\newcount\sectna  %section with analysis (IV)
\newcount\VFE
\newcount\VFV
\newcount\PLT
\newcount\PLN
\newcount\KWW
\newcount\VS
\newcount\FE
\newcount\RYF
\newcount\PYC
\newcount\TU
\newcount\FT
\newcount\FB
\newcount\LA
\newcount\EMX
\newcount\EMY
\newcount\IE
\newcount\NS
\newcount\ET
\newcount\ES
\newcount\QE
\newcount\EC
\newcount\ST
\newcount\LU
\newcount\TU
% other counts
\newcount\lpnn
\newcount\lpno
%DEFINITIONS
\def\r{\rho}
\def\l{\lambda}

\def\vr{{\bf r}}
\def\vg{{\bf g}}
\def\vj{{\bf j}}
\def\vx{{\bf x}}
\def\vq{{\bf q}}
\def\dl{\nabla}
\def\tp{\tau^\prime}
%other stuff

\def\table{\topinsert
\vskip -2pt plus 4pt minus 4pt
\baselineskip=2truept \parindent=2.5truecm
\endinsert}

\newcount\rfno  \rfno=0
\def\ref#1  {\advance\rfno by1 {$^{\number\rfno}$}
\setbox\number\rfno=\hbox{#1}}
\def\incref#1  {\advance\rfno by1 \setbox\number\rfno=\hbox{#1}}

\def\reference {\par{\leftline{\bf REFERENCES}}\lpno=\number\rfno
\advance\lpno by 1 \lpnn=1 \loop \item{\number\lpnn .} {\unhbox\number\lpnn}
\advance\lpnn by 1
\ifnum \number\lpnn <\number\lpno \repeat}
\def\no {\number}
\def\appl {{\raise2pt\hbox{$\textstyle{<}$}}\hskip -
7pt{\lower2pt \hbox{$\textstyle\sim$}}}
%
%   Count Footnotes
%
\newcount\notenumber
\def\clearnotenumber{\notenumber=0}
\def\fnote{\advance\notenumber by1 $^{\the\notenumber}$}
\clearnotenumber
%
%   Count Equation Numbers
%
\newcount\eqnumber
\newcount\secno \secno=0
\def\newsection{\global\advance\secno by1 \eqnumber=0}
\def\eqnch{\global\advance\eqnumber by1 \eqno({\the\secno}.{\the\eqnumber})}
\def\eqna{\global\advance\eqnumber by1 ({\the\secno}.{\the\eqnumber}a)}
\def\eqnb{({\the\secno}.{\the\eqnumber}b)}
\def\eqnc{({\the\secno}.{\the\eqnumber}c)}
\baselineskip=22pt plus 1pt

\parindent 30pt
\font\titlefnt=cmbx10 scaled\magstep3
\font\auf=cmr10 scaled \magstep1

\hbox{}
\centerline{\titlefnt Nonlinear hydrodynamics of a hard sphere  }
\centerline{\titlefnt fluid near the glass transition}
\vskip .3truein
\centerline{\auf Lisa M. Lust and Oriol T. Valls }
\vskip .2truein
\centerline {\auf School of Physics and Astronomy}
\centerline {\auf and Minnesota Supercomputer Institute,}
\centerline {\auf University of Minnesota,}
\centerline {\auf Minneapolis, Minnesota 55455-0149}
\centerline {\auf and}
\centerline {\auf Chandan Dasgupta}
\vskip .2truein
\centerline {\auf Department of Physics,}
\centerline {\auf Indian Institute of Science,}
\centerline {\auf Bangalore 560012, India}
\vfill
\eject
\centerline{\bf Abstract}
\medskip
We conduct a numerical study of the  dynamic behavior of a dense hard sphere
fluid by deriving and integrating a set of Langevin equations.
The statics of the system is described by a free energy functional of the
Ramakrishnan-Yussouff form. We find that the system exhibits glassy behavior
as evidenced through stretched exponential decay and two-stage relaxation of
the density correlation function. The characteristic times grow with
increasing density according to the Vogel-Fulcher law. The wavenumber
dependence of the kinetics is extensively explored. The connection of our
results with experiment, mode coupling theory, and molecular dynamics
results is discussed.
\vskip .1truein
\noindent

\noindent
1992 PACS numbers:  64.70Pf, 61.20Ja, 61.20Lc
\vfill
\eject

\centerline{I. INTRODUCTION}
\newsection {\global\sectni=\no\secno}

Despite extensive experimental, numerical and theoretical investigations
\ref{J. Jackle,
Rep. Prog. Phys. {\bf 49}, 171 (1986).}{\global\JA=\number\rfno} $^,$ \ref{C.
A. Angell, J. Phys. Chem. Solids, {\bf 49}, 863, (1988).}
{\global\AN=\number\rfno} over
several decades, the present understanding of the slow
non-exponential dynamics of dense liquids near the glass transition
remains incomplete. When a liquid is cooled rapidly enough to
temperatures below the equilibrium freezing temperature,
crystallization is bypassed and  the system undergoes a transition
into an amorphous solid state called a glass. The characteristic relaxation
time $t^*$ , as reflected in a large number of
experimentally measured quantities  such as viscosity and dielectric
relaxation, grows rapidly in the supercooled state as the temperature
is decreased or the density increased.
The glass transition temperature (or alternatively, density) $T_g$ ($\r_g$) is
conventionally defined as the temperature (density) where the
viscosity reaches a value of $10^{13}$ poise.

In recent years, considerable progress in the development of a
theoretical and experimental
understanding of the phenomena associated with glass formation and the
glass transition
has been achieved. In particular, the so-called mode coupling (MC) theories
of the glass \incref{For reviews of mode coupling theories, see W.
G\"{o}tze and L. Sjogren in {\it Dynamics of Disordered Materials}, eds.
D. Richter, A. J. Dianoux, W. Petry and J. Teixeira (Springer,
Berlin, 1989); B. Kim and G. F. Mazenko, Adv. Chem. Phys. {\bf 78},
129 (1990). } {\global\MC=\number\rfno}\incref{E.
Leutheusser, Phys. Rev. {\bf A29}, 863 (1984).}
{\global\LE=\number\rfno}\incref{U.
Bengtzelius, W. G\"{o}tze and A. Sjolander, J. Phys. {\bf C17}, 5915
(1984).}{\global\BGS=\number\rfno}
\incref{W.G. G\"{o}tze, Z.Phys. {\bf 60}, 195, (1985) .}
{\global\GO=\number\rfno}
\incref{W. G\"{o}tze and L. Sjogren, Z. Phys. {\bf B65},
415 (1987); J. Phys. C{\bf 21}, 3407 (1988); J. Phys. Cond. Matt.
{\bf 1}, 4183 (1989).} {\global\GS=\number\rfno}
\incref{T.R. Kirkpatrick, Phys.Rev. {\bf A31}, 939, (1985) .}
{\global\KR=\number\rfno}
\incref{S. P. Das and G. F. Mazenko, Phys. Rev. {\bf A34},
2265 (1986).} {\global\DM=\number\rfno} \incref{S. P. Das, Phys. Rev.
{\bf A36}, 211 (1987).} {\global\DAS=\number\rfno}
transition $^{ {\no\MC}-{\no\DAS} }$ have led to a framework for understanding
and interpreting many experimental results. In MC theories,
the slowing down of the dynamics near the glass transition is
attributed to a nonlinear feedback mechanism arising
from correlations of density fluctuations in the liquid.
The MC equations for the dynamics near the glass transition were originally
derived $^{ {\no\LE},{\no\BGS} }$ from the kinetic theory of dense fluids
and were later generalized and extended by G\"{o}tze and co-workers
$^{ {\no\GO},{\no\GS} }$.
In the original version of MC
theories $^{ {\no\LE},{\no\BGS} }$, the characteristic time scales of
the liquid are predicted to exhibit a power law divergence
at an `ideal glass transition' or crossover temperature $T_c$, higher
than $T_g$.  However, this
divergence is not found experimentally: the power law form breaks
down, and relaxation times at $T_c$ are typically of order $10^{-8}$s.
More recent
calculations $^{ {\no\DM},{\no\DAS} }$ have uncovered a cutoff
mechanism which is supposed to round off the predicted divergence and
to restore ergodicity over a much longer time scale.
Most of the existing MC studies of the glass transition do not take into
account the equilibrium short-range structure of the dense liquid. As a
result, these theories do not lead to any prediction for the wavenumber
dependence of the relaxation. Some attempts towards the incorporation
of information about liquid structure in the MC formalism have been made
recently $^{ {\no\KR},{\no\DAS} }$. MC theories
$^{{\no\MC},{\no\GS}}$
and experimental information
can be combined in a consistent description of the decay of $S(\vq,t)$,
the spatial Fourier transform of the dynamic density correlation function,
at sufficiently low temperatures or high densities, when the decay times
have grown very large. The decay
according to this scheme takes place in a succession of several
regimes: after a fast decay in times of order of the inverse phonon
frequency, a first slow decay occurs which MC theories predict to be an
inverse power law in time. This is called the $\beta$-relaxation regime.
Evidence for power-law decay of correlations in the $\beta$
regime is provided by light
\ref{N. Tao, G. Li, and H.Z. Cummins, Phys. Rev. Lett {\bf 60}, 1334,
(1990).} and neutron scattering  experiments
\ref{See for example, B. Frick, B. Farago, and D. Richter,
Phys. Rev. Lett. {\bf 64}, 2921, (1990); W. Petry {\it et al}, Z. Phys.
{\bf B83}, 175, (1990).} .This decay is to a nonzero value
(an apparent
nonergodic phase) from which the system eventually moves away leading to
the primary or $\alpha$-relaxation regime.
The relaxation in the
$\alpha$ regime is found to follow the so-called Kohlrausch -
Williams - Watts `stretched exponential' form\ref{G. Williams and
D. C. Watts, Trans. Faraday Soc. {\bf 66}, 80 (1970).}
{\global\KW=\number\rfno}.
Both the duration of
the $\beta$ relaxation and the time scale  of the
stretched exponential decay in the $\alpha$ regime are found to
increase sharply as the `glass transition' is approached. In some
cases, the $\beta$ and the $\alpha$ regimes are separated by a region
of  so-called von Schweidler relaxation which has a power law form. The
stretched exponential behavior in the $\alpha$
regime and the von Schweidler relaxation are seen in
dielectric measurements \ref{P.K. Dixon, L.Wu, S.R. Nagel, B.D. Williams,
 and J.P.Carini, Phys. Rev. Lett. {\bf 66}, 960, (1991).} and in light
\ref{I.C. Halalay and K.A. Nelson, Phys. Rev. Lett. {\bf 69}, 636, (1992).}
and neutron scattering \ref{ J. Colmenero, A. Alegria, A. Arbe, and B. Frick,
Phys. Rev. Lett. {\bf 69}, 478, (1992). See also other neutron work cited
above.} experiments. The parameters which describe the decay of
$S(\vq,t)$ are known to be $q$-dependent, but the nature of this
dependence has not been studied in detail.
Thus, the MC theories provide a qualitative
understanding of a number of experimentally observed features of
glassy relaxation. However, some of the detailed MC predictions are
not in agreement with experiments \ref{B. Kim and G.F. Mazenko, Phys.
Rev. {\bf A45}, 2393, (1992).} {\global\KM=\number\rfno}
and the MC description clearly fails
to account for the behavior observed at temperatures close to and
lower than $T_c$. It is generally believed $^{\no\MC}$ that this
failure arises from the fact that MC theories do not take into
account activated processes involving transitions between different
local minima of the free energy which are supposed to develop as the
temperature is lowered below the equilibrium freezing temperature.

In contrast to MC theories which portray the glass transition as being
purely dynamic in nature, there have been a
number of attempts \ref{T. R. Kirkpatrick, D. Thirumalai and P. G.
Wolynes, Phys. Rev. {\bf A40}, 2 (1989).} $^-$ \incref{D. L. Stein and
R. G. Palmer, Phys. Rev. {\bf B38}, 12035 (1988).} \incref{J. P. Sethna,
Europhys. Lett. {\bf 6}, 529 (1989).} \ref{J. P. Sethna, J. D. Shore
and M. Huang, Phys. Rev. {\bf B44}, 4943 (1991).} to develop a
`thermodynamic' theory in which some of the
interesting behavior observed near the glass transition is
attributed to an underlying continuous phase transition. These
attempts have been motivated by the fact that the observed growth of
the relaxation time in so-called `fragile' liquids $^{\no\AN}$ is
well-described  by the Vogel-Fulcher law \ref{H. Vogel, Z. Phys {\bf 22},
645 (1921); G. S. Fulcher, J. Am. Ceram. Soc. {\bf 8}, 339 (1925).}
{\global\VF=\number\rfno}:
$$ t^* = \tau_s e^{a T_0\over{T-T_0} } \eqnch$$
{\global\VFE=\number\eqnumber}
\noindent where $\tau_s$ is a characteristic microscopic time, $a$ is
a dimensionless constant and $T_0$ is a characteristic temperature
which is found to be well below the conventional glass transition
temperature $T_g$.
This law  suggests the possibility of a
so-called `thermodynamic glass transition' that would take place at the
temperature $T_0$ if thermodynamic equilibrium \ref{Here, and elsewhere
in the paper, the term
`thermodynamic equilibrium' means
equilibrium over a restricted phase space that excludes the
crystalline state .}
could be maintained down to this temperature. Such a transition is
also suggested by the fact $^{\no\AN}$ that the Kauzmann
temperature \ref{W. Kauzmann, Chem. Rev. {\bf 48}, 219 (1948).} at
which the entropy difference between the supercooled liquid and the
crystalline solid extrapolates to zero is close to $T_0$. Since in
practice the system falls out of equilibrium at temperatures close
to $T_g$, a direct experimental test of
the existence of such a transition is not possible. Also, any
calculation that explicitly demonstrates the existence of such a
transition in a physically realistic system is not yet available.
Thus, the `thermodynamic glass transition'
scenario remains essentially speculative. In this approach, the growth of
the relaxation time is attributed to a growing correlation length which
would diverge at the ideal glass transition.
Two recent numerical studies\ref{C. Dasgupta, A. V. Indrani, S.
Ramaswamy and M. K. Phani, Europhys. Lett. {\bf 15}, 307 (1991).} $^,$
\ref{R. M. Ernst, S. R. Nagel and G. S. Grest, Phys. Rev. {\bf B43},
8070 (1991).}  which looked for such a growing correlation length did
not find any evidence for its existence. A recent experiment
\ref{E.W. Fisher, E. Donth, and W. Stephen, Phys. Rev. Lett.
{\bf 68}, 2344, (1992). } , on the other hand, has presented evidence for the
existence of a length scale
that grows as the temperature is lowered below the crossover
temperature $T_c$.

The  static and dynamic properties of dense liquids have also been
studied extensively by molecular dynamics (MD) simulations
\ref{L. V. Woodcock and C. A. Angell, Phys. Rev. Lett. {\bf 47}, 1129
(1981).} {\global\WA=\number\rfno}$^-$\incref{J. J.
Ullo and S. Yip, Phys. Rev. Lett. {\bf 54}, 1509, (1985); Phys. Rev.
{\bf A39}, 5877 (1989).} \incref{ J.J. Ullo and S.
Yip, Chem. Phys. {\bf 149}, 221 (1990).} {\global\UY=\no\rfno} \incref{G.
F. Signorini, J. -L. Barrat and M. W. Klein, J. Chem. Phys. {\bf 92},
1294 (1990).} {\global\SIG=\no\rfno}
\ref{J. -L. Barrat, J. -N. Roux and J. -P. Hansen, Chem.
Phys. {\bf 149}, 197 (1990).}
{\global\BRH=\no\rfno}. The very nature of the simulation
method restricts such studies to simple model systems and to time
scales which are rather short (of the order of $10^{-9}$ sec).
In spite of these limitations, MD simulations have produced a number
of interesting results which are in qualitative agreement with
predictions of MC theories and results of experiments (such as those
using the neutron spin-echo technique \ref{F. Mezei, W. Knaak and B.
Farago, Phys. Rev. Lett {\bf 58}, 571 (1987); Physica Scr. {\bf 19},
 363 (1987).} ) which have time
scales comparable to those of the simulations. This observation
leads to the interesting and useful conclusion that a study of simple
model systems may be sufficient for understanding the basic physics
of the glass transition.

It is evident from this brief survey of the current status of the
glass transition problem that an obvious need exists for the
development of new analytic and numerical methods which may address
some of the outstanding issues related to this problem. In this
paper, we describe the results obtained from the application of
a new numerical method to a study of the
dynamic behavior of a dense hard-sphere liquid near the glass
transition. The method we use consists of direct numerical
integration of a set of Langevin equations which describe the
nonlinear fluctuating hydrodynamics (NFH) \ref{NFH as formulated here
was first used in the study of fluids by O.T. Valls and G.F. Mazenko,
Phys. Rev. {\bf B44}, 2596, (1991). For a review of the formalism, see
B. Kim and G. F. Mazenko,
J. Stat. Phys. {\bf 64}, 631 (1991).} of the system. Information about
the static structure of the liquid is incorporated in the Langevin
equations through  a free-energy functional which has a form
suggested by Ramakrishnan and Yussouff (RY).\ref{T. V. Ramakrishnan
and M. Yussouff, Phys. Rev. {\bf B19}, 2275, (1979); D. W. Oxtoby in {\it
Liquids, Freezing and the Glass Transition, Lecture Notes of the Les
Houches Summer School, 1989}, ed. J. -P. Hansen, D. Levesque and J.
Zinn-Justin (Elsevier, New York, 1990).}
{\global\RY=\number\rfno} Recently, it has been shown \ref{C.
Dasgupta, Europhys. Lett. {\bf 20}, 131 (1992); C. Dasgupta and S.
Ramaswamy, Physica A {\bf 186}, 314 (1992).} {\global\CD=\number\rfno}
that the RY free energy functional provides a correct mean-field
description of the statics of the glass transition in this system.
In that work, a numerical procedure was used to locate local minima
of a discretized version of the RY free energy appropriate for the
hard sphere system. A large number of glassy local minima with
inhomogeneous but aperiodic density distribution were found to appear
as the average density was increased above the value at which
equilibrium crystallization takes place. At higher densities, the
free energies of these minima were found to drop below that of the minimum
representing the uniform liquid signaling a mean-field glass
transition. The success of the RY free energy functional in providing
a correct description of the statics of the glass transition of the hard
sphere system suggests that a good starting point for a study of the
dynamics of this system would be obtained by incorporating this free
energy in the appropriate NFH equations.

A number of important issues
are addressed in our study of the dynamics. By comparing the results of our
calculations with existing MD results $^{\no\WA}$ on the same system,
we are able to test the validity of the NFH description which is cast
in terms of
coarse-grained number and current density variables instead of the
coordinates and momenta of individual particles. The correctness of the
NFH equations we use, although usually taken for granted, is not
obvious in view of the fact that the hydrodynamic terms in these
equations describe the physics at relatively long length scales,
whereas the terms arising from the free energy functional involve
length scales of the order of (or smaller than)
the interparticle spacing.
In the RY free energy functional,
information about the microscopic interactions is incorporated in the
form of the Ornstein-Zernike direct pair correlation function \ref{J.
P. Hansen and I. R. McDonald, {\it Theory of Simple Liquids},
(Academic Press, London, 1986).} {\global\PY=\number\rfno} of the
liquid. This appears to be adequate for a correct description of the
statics of the freezing of the liquid into both  crystalline
$^{\no\RY}$ and glassy $^{\no\CD}$ states. One of the questions we
address in the present study is whether this is also sufficient for a
correct description of the dynamic behavior. Two numerical
studies of NFH equations describing the dynamics of dense liquids have been
reported recently \ref{O. T. Valls and G. F. Mazenko, Phys. Rev. {\bf
B44}, 2556 (1991).} $^,$ \ref{O. T. Valls and G. F. Mazenko, Phys.
Rev. A, to appear.} {\global\VM=\number\rfno}. The main difference
between these studies and the present one is that the equilibrium
structure of the liquid was treated only in an approximate way in
the earlier calculations. Consequently the results of these
calculations could not be compared directly with MD data. The results
of these calculations did not exhibit some of the features (such
as two-regime decay of correlations) generally associated with
glassy dynamics, although non-exponential decay of correlations
often associated with its onset was found.
In Ref.~$({\no\VM})$, it was suggested that this failure arises from the
approximations made in the treatment of liquid structure. The present
study provides the opportunity
to check whether this explanation is correct and
to investigate in general the role of static structure in the dynamics
of a dense liquid.
Further,
perturbative treatments of the NFH equations similar to the ones we use are
known
$^{{\no\DM},{\no\DAS}}$  to lead to results which are very close to those
obtained from the MC approach.
Apart from numerical
errors arising from spatial discretization and the integration
procedure, our treatment of these equations is exact. In particular,
our numerical solution of these equations
is obviously nonperturbative, therefore, a comparison of the
results of our calculation with MC predictions provides a way to test
the validity of some of the approximations made in the analytic studies.
Since our calculation takes full account of the short range structure of
the liquid, it has the most direct connection with the MC studies of
Kirkpatrick $^{\no\KR}$ and Das $^{\no\DAS}$.
Finally, by monitoring which minima of the free energy are visited
during the time evolution of the system we are able to determine
whether the observed dynamic behavior arises from nonlinearities of
density fluctuations in the liquid or from transitions among
different glassy minima of the free energy. It is not possible to
distinguish between the effects of these two kinds of processes
in conventional MD simulations.

We have studied the time-decay of $S(q,t)$, the spherically averaged
spatial Fourier transform of the time-dependent density correlation
function of the hard-sphere liquid, for different values of
wavevector $q$ and
for a number of values of the reduced density $n^*$ ($n^*\equiv \r_0
\sigma^3$, where $\r_0$ is the average number density and $\sigma$ is the hard
sphere diameter) in the range $0.75 - 0.93$. It was found in
Ref.~(${\no\CD}$) that a hard sphere system described by
a discretized version of the RY free energy
exhibits a crystallization transition near
$n^*=0.83$.
Since the present calculation uses the
same discretized free energy as that of  Ref.(${\no\CD}$), the system
may be considered to be in the `supercooled' regime for a large part
of the density range considered by us.
The numerical efficiency of the Langevin dynamics arising from the
use of coarse grained variables enables us to verify that the statics of the
system remain stationary throughout the long time intervals that we
consider.  The observed dynamic behavior
described later in detail exhibits a number of
characteristic glassy features. These include stretched exponential
decay of correlations, two-stage relaxation,
and Vogel-Fulcher growth of relaxation times. Our results are in
agreement with existing MD data$^{\no\WA}$ on the dynamics
of the hard-sphere liquid and in qualitative agreement with other MD results
obtained for similar but different systems. This observation establishes the
correctness of the NFH description used in this work
and also demonstrates that
the RY free energy contains the essential physics of the dynamics of
this system. Our calculations also reproduce qualitatively a number of
predictions of MC theories, however, we find significant
deviations from some of
the quantitative MC predictions \ref{J. L. Barrat, W. G\"{o}tze and A.
Latz, J. Phys. Condens. Matt. {\bf 1}, 7163 (1989).} for the glassy
kinetics of the hard-sphere system. Our study indicates that
the onset of glassy features in the decay of $S(q,t)$ occurs at
relatively lower densities for wavevectors close to the
first and second peaks in the static structure factor.
This result clearly
illustrates the important role played by the equilibrium structure in
the long-time dynamics of the liquid. The observed $q$-dependence of
the decay of $S(q,t)$ also suggests that the glassy behavior sets in
earlier (at lower densities) at shorter length scales. Finally,
the system is found to fluctuate about the liquid-state
minimum of the mean field free energy in all our simulations.

The rest of the paper is organized as follows. Section II contains a
description of the model considered by us. We first define the model,
discuss its statics and  derive the appropriate NFH equations.
By defining appropriate units of length, time and mass, these equations
are then written in dimensionless form.
The method used by us to integrate these
equations forward in time is described in Section III. We also
discuss in this Section the physical quantities  measured, the data
collection procedure, tests of equilibration and other related
matters.  The results obtained from the numerical work are described
in detail in Section IV. We also compare and contrast our results
with those obtained from MD simulations of the hard-sphere and
similar systems and the predictions of MC theories.
\bigskip

\centerline{II. THE MODEL}
\newsection {\global\sectnf=\no\secno}

\bigskip

We have explained in the Introduction that we will analyze the numerical
solution of a set of Langevin equations appropriate to a dense hard-sphere
fluid. We begin here by discussing the statics of the model. These are
given in terms of a free energy which is a functional of the two fields in
the problem: the density field $\r(\vr,t)$ and the current field
$\vg(\vr,t)$. This free energy has two terms:
$$ F_{Tot}[\r,\vg]=(m_0/2) \int{d \vr {g^2 \over \r_0}} + F[\r] \eqnch$$
{\global\FE=\number\eqnumber}
\noindent where $F[\r]$ is of the RY form $^{\no\RY}$:
$$F[\r]=F_l[\r_0] +
k_B T \left(\int{d \vr(\r \log(\r/ \r_0)-\delta \r)}-(1/2)
\int{d \vr d\vr^\prime C(\vr-\vr^\prime) \delta \r \delta \r^\prime} \right)
\eqnch$$ {\global\RYF=\number\eqnumber}
\noindent In $(\no\sectnf.\no\FE)$  and $(\no\sectnf.\no\RYF)$ $\r$ is
the number
density field and $\delta \r \equiv \r-\r_0$ the deviation of that field from
its average value $\r_0$. $F_l$ is the free energy of the uniform liquid,
$T$ is the temperature and $k_B$ the Boltzmann
constant. In the first equation $m_0$ denotes the mass of a hard sphere.
Finally, $C(\vr-\vr^\prime)$ is the direct correlation function. The inclusion
of this function in the free energy ensures that upon linearization of the
logarithm in the first term on the right hand side of $(\no\sectnf.\no\RYF)$
one obtains the usual expression for the static structure factor of a
simple fluid in terms of
$C$. For hard spheres a simple expression for $C(\vr-\vr^\prime)$ can be
obtained in the Percus-Yevick
approximation $^{\no\PY}$:
$$\eqalignno{C(\xi)&=-\l_1-6 \eta_f \l_2 \xi -(1/2)\eta_f \l_1 \xi^2 ;\
\ \ \xi<1 &\eqna\cr
C(\xi)&=0\ \ ;\ \ \xi>1 &\eqnb\cr} $$ {\global\PYC=\number\eqnumber}
\noindent where $\xi \equiv |\vr-\vr^\prime|/ \sigma$,
$\eta_f$ is the packing fraction:
$$\eta_f=(\pi/6)\r_0 \sigma^3 \equiv (\pi/6) \ n^* \eqnch $$
\noindent and:
$$\l_1=(1+2\eta_f)^2/(1-\eta_f)^4 \eqnch$$
$$\l_2=-(1+\eta_f/2)^2/(1-\eta_f)^4 \eqnch$$
We have written $\r_0$ in the denominator of the $g^2$ term in
Eq.$(\no\sectnf.\no\FE)$, rather than $\r$ as in
 Ref.~({\no\VM}), so that the full density dependence of the free energy is
given by the RY form. As pointed out in
\incref{O.~T. Valls and G.~F. Mazenko, Phys Rev {\bf B38}, 11643, (1988).}
{\global\VMO=\no\rfno} Ref.~({\no\VMO}), if we used $\r$ we would obtain
upon functional integration with respect to the gaussian variable $g^2$
a $\log (\r)$ contribution involving density fluctuations associated with the
kinetic energy. These are already included in
Eq. $(\no\sectnf.\no\RYF)$.

It is convenient to choose units at this point, which will simplify subsequent
expressions. We take $m_0$ as the unit of mass. We choose also a unit of
length,
$h$, which we shall later identify with the lattice constant of the
computational lattice. For the unit of time we choose $t_0$ such that:
$$t_0=h/c \eqnch$$ {\global\TU=\number\eqnumber}
\noindent where $c$ is the speed of sound. This choice of the unit of time is
motivated as follows: the characteristic phonon time for the system is $t_p=
1/(cq)$ where $q$ is a wavevector. Hence $t_p/t_0=1/(q h)$. We shall choose
below the ratio $\sigma/h$ so that $ q h$ is of order unity in the wavevector
range of interest. Hence the choice $(\no\sectnf.\no\TU)$ corresponds to a
characteristic time of order $t_p$. The same choice was taken in Ref.~({\no
\VM}).

We then define the dimensionless quantities:
$$\vx=\vr/h \eqnch$$
$$n=\r h^3 \eqnch$$
$$\vj=\vg h^3/c \eqnch$$
\noindent and introduce also the dimensionless free energy $F[n,\vj]$ in terms
of the above quantities:
$$F[n,\vj]=(1/2)\int{d \vx{ j^2\over n_0}}+ F_n[n] \eqnch$$
{\global\FT=\number\eqnumber}
$$F_n[n]=F_{ln}[n_0] + K \left( \int{d \vx (n \log (n/n_0)-\delta
n)}-(1/2)\int{
d\vx d\vx^\prime C(\vx-\vx^\prime) \delta n \delta n^\prime} \right)
\eqnch$$ {\global\FB=\number\eqnumber}
\noindent where
$$K={k_B T\over{m_0 c^2}}  \eqnch $$
\noindent The quantity K is for hard spheres a function of only the density.
This follows from:
$$c^2={1\over{m_0 \r_0 \kappa}} \eqnch$$
\noindent where the compressibility $\kappa$ is related to the structure factor
$S(\vq)$ by $k_B T \r_0 \kappa=S(\vq=0)$. The latter quantity can
straightforwardly be calculated from $(\no\sectnf.\no\PYC)$ and one obtains:
$$K=\left(1+(4 \pi/3)n^*(\l_1+(9/2) \eta_f \l_2 +(\eta_f/4)\l_1)\right)^{-1}
\eqnch$$
\noindent which indeed depends on the density only. The quantities $\l_1,\l_2,
\eta_f,$ and $n^*$ were defined above.

We quote here the relation between our unit of time $t_0$ and the Enskog
mean collision time $t_c$ commonly used in MD work \ref{W.E. Alley, B.J. Alder,
and S. Yip, Phys.Rev. {\bf A27}, 3174, (1981)} {\global\AAY=\no\rfno}
on hard spheres. One easily finds that:
$$t_c^{-1}=4 \sqrt{\pi}n^* g(\sigma) \sqrt{K} (h/\sigma) t_0^{-1}\eqnch$$
{\global\TU=\no\eqnumber}
\noindent where $g(\sigma)$ is the pair correlation function at contact\ref
{J. P. Boon and S. Yip, {\it Molecular Hydrodynamics},
McGraw-Hill, New York, (1990).
x Eq. $2.4.40$.}{\global\BYP=\no\rfno} .
The static properties of a system described by the RY free energy, and its
relation to glassy behavior have been discussed in the literature
$^{{\no\CD},{\no\RY}}$ and in the Introduction.
We turn now to the dynamics. The hydrodynamic variables are of course the
fields $n$ and $\vj$. We denote, following the ususal notation, this set of
Langevin variables as $\{\psi_\alpha\}\equiv \{n(\vx,t),\vj(\vx,t)\}$.
Formally,
the Langevin equations can be written as $^{\no\DM}$
$^,$\ref{J. E. Farrell and O. T. Valls, Phys. Rev.
{\bf B40}, 7027, (1989).} {\global\FV=\no\rfno}:
$${ \partial \psi_\alpha\over{\partial t}}= V_\alpha[\psi]-\sum_\beta \Gamma_{
\alpha,\beta} {\delta F\over \delta \psi_\beta} + \Theta_\alpha \eqnch$$
{\global\LA=\number\eqnumber}
\noindent where $\Gamma$ is the matrix of transport coefficients, $\Theta$
are the gaussian noise fields, and the $V_\alpha$ are the streaming velocities,
which incorporate the nondissipative part of the equations of motion.
Proceeding
as in Refs.~({\no\DM}) and ({\no\FV) one obtains:
$$V_n=-\sum_j \dl_j[n(\vx){\delta F\over {\delta j_j(\vx)}}] \eqnch $$
$$V_{j_i}=-n(\vx) \dl_i {{\delta F_n} \over {\delta n(\vx)}} - \sum_j \dl_j
{j_i(\vx)j_j(\vx)\over n_0}-\sum_jj_j(\vx)\dl_i{j_j(\vx)\over n_0} \eqnch $$
\noindent where $F$ and $F_n$ are defined in  Eqns. $({\no\sectnf}.\no\FT)
$ and $(\no\sectnf.\no\FB)$. Following then on precisely as in Ref.~({\no\VM})
we obtain, after straightforward algebra the equations of motion:
$${ \partial n(\vx,t)\over{\partial t}} +(1/n_0) \dl (n \vj)=0 \eqnch $$
 {\global\EMX=\number\eqnumber}
\noindent and:
$${\partial j_i\over{\partial t}}=-n \dl_i {\delta F_n \over{\delta n}}-(1/n_0)
\sum_j\dl_j(j_ij_j)-(1/n_0)\sum_jj_j\dl_i j_j+ (1/n_0) \eta \dl^2 j_i +
\Theta_i  \eqnch $$  {\global\EMY=\number\eqnumber}
where $\eta$ is the bare shear viscosity in our units \ref{We have set the
combination $(\zeta+\eta/3)$, where $\zeta$ is the bare bulk viscosity, to
zero, mainly for reasons of simplicity in generating the noise
correlations.} . The noise fields
$\Theta_i(\vx,t)$ satisfy the second fluctuation-dissipation theorem in the
form:
$$<\Theta_i(\vx,t)\Theta_j(\vx^\prime,t^\prime)>=-2 K \lambda \eta n_0 \delta_
{i,j} \dl^2 \delta(\vr-\vr^\prime) \delta(t-t^\prime) . \eqnch $$ {\global\NS=
\number\eqnumber}

We recall that in these equations, and in the remainder of the paper, the time
is measured in units as given in $(\no\sectnf.\no\TU)$. In
$(\no\sectnf.\no\NS)$ the angular brackets denote the thermodynamic average.
The quantity $\l$ is a dimensionless measure of the equilibrium fluctuations.
The average value $n_0$ is related to $n^*=n \sigma^3$ through
$n_0=n^*(h/\sigma)^3$. For hard spheres, \ref{J.P. Boon and S. Yip,
{\it Molecular Hydrodynamics}, McGraw-Hill, New York, (1990), Eq. $ 6.7.21$.}
{\global\BY=\no\rfno} one can write $\eta$
in terms of $K$ and the density in the form:
$$ \eta =(h/ \sigma)^{2} \sqrt{K} [5/(16\sqrt{\pi})][g(\sigma)^{-1}
+ 0.8(2\pi n^*/3) +0.761(2\pi n^*/3)^2 g(\sigma)]
   \eqnch$$ {\global\ET=\number\eqnumber}
The equations of motion
$(\no\sectnf.\no\EMX)$ and $(\no\sectnf.\no\EMY)$ are
slightly more complicated than the corresponding equations for
the model in Ref.~({\no\VM}). There is an additional convection term in the
second equation and some additional factors of $n/n_0$. These
complications can be directly traced down to the
different density dependence of the first term in $(\no\sectnf.\no\FT)$
as discussed above, that is, to the fact that a kinetic energy contribution
is now included in the first term of $F_n$. These differences then arise
because of the different form of  $\delta F/\delta\vj$ which
results in more complicated expressions for the streaming velocities. Our
expressions correctly reduce to the continuity and Navier-Stokes equations
in the linear limit. In general, one can interpret $(\no\sectnf.\no\EMX)$
as the continuity equation if the field $\vg$  is thought of as $\rho_0$
times the velocity. The current density is then not given simply by $\vg$,
however. These questions do not affect, obviously, the validity of the
conclusions obtained from this model.

A salient feature of Eq.$(\no\sectnf.\no\EMY)$
of considerable importance is that the term in
${\delta F_n}/{\delta n}$ which involves the direct correlation function
$C$ is an integral over space with range $\sigma$. Thus, we have to solve in
this case not merely a set of partial differential equations with stochastic
terms but as far as the spatial dependence is concerned an
integrodifferential equation. This complication makes this a difficult problem
to solve numerically. The methods used will be explained below.

\bigskip

\centerline{III. METHODS}
\newsection {\global\sectnr=\no\secno}

\bigskip

In this Section we discuss the methods we use to solve our equations of motion,
the physical quantities we focus on, the data collection methods and
statistics,
the range of parameter values we have studied, and related matters.

Our objective is to study
the dynamic correlations of our system as defined below. In order
to do so we solve numerically Eqns.~$(\no\sectnf.\no\EMX)$ and
$(\no\sectnf.\no\EMY)$ on a three dimensional cubic
lattice of size $N^3$. The main complication we must consider is the spatial
integral involving the direct correlation function $C(r)$.
For each integration step this integral must be done $N^3$ times since it is
computed over a sphere with its origin at each one of
the lattice sites.
To avoid repeated calculations we create, in the initialization of our program,
a table listing for each lattice site the location of all neighboring
sites to be
integrated over and their corresponding values of $ C(r)$. Additionally,
since the sphere is imbedded on a coarse discrete lattice we use a finer mesh
than defined on the original lattice to improve the accuracy of the
integration.
The procedures employed to integrate the remaining set of differential
equations  over time and generate the gaussian noise are
identical to  those used in Ref.~({\no\VM}) and references cited there.

We will focus our analysis on the time dependence of the dynamic
structure factor $S(\vq,t)$:

$$S(\vq,t) = \int{d^3x e^{i\vq\cdot(\vx-\vx^\prime)}<\delta n(\vx,0)
\delta n(\vx^\prime,t)>} \eqnch$$ {\global\IE=\no\eqnumber}
\noindent Specifically, we will consider the angular average of $S(\vq,t)$. On
a cubic lattice, it is appropriate to define the effective length of $\vq$,
$q$,
as:
$$ q^2=2(3-\cos q_x -\cos q_y -\cos q_z), \eqnch$$
{\global\QE=\no\eqnumber}
and we perform angular averages of $\vq$-dependent quantities by averaging
over values of $\vq$ in the first Brillouin zone,
in a spherical shell of mean radius (as given by $q$)
corresponding to that of the vector $(\pi Q/N,0,0)$ and thickness
$\pi/N$. The value
of $Q$ ranges from 1 to approximately $3^{1/2}N$, although only a smaller
range is free of finite size effects. We will, for simplicity of notation,
denote quantities averaged in this way by simply dropping the vector symbol
from the wavevector argument: $S(q,t)$, and often we will indicate the values
of $q$ by the `shell number' $Q$.

Next we turn to our choices for parameter values. The correlation functions
that we are interested in are spatially short ranged. It is therefore not
necessary to use extremely large lattice sizes. The results presented were
obtained using $N=15$. As in Ref.~({\no\VM}), this proved to be adequate. We
checked that finite size effects do not affect the dynamics in the wavevector
region for which results are presented here, ($5<Q<15$, see Section IV)
by performing a portion of the calculations (with reduced
statistics) at $N=25$.
Our choice of the ratio $\sigma/h$ which fixes the
length scale for the problem, is dictated by two concerns.
The first is that we wish to be able to study the dependence of
the dynamics on wavevector in the main region of interest from the point of
view of the static structure factor $S(q)\equiv S(q,t=0)$. Thus, we wish to
choose our
unit of length so that the main peak in $S(q)$ falls in the middle part of
the range of wavevectors within the first Brillouin zone of the computational
lattice.  Secondly, to avoid crystallization at the higher densities studied,
we have found that
we need N and $\sigma$ to be incommensurate.
Selecting $\sigma/h=4.6$ leaves $q_{max}$ well away from the zone edge for all
densities, near the $Q=8$ shell, and is clearly incommensurate with N.
The density range we have investigated includes $n^*=0.5$, $0.75 \le n^*\le
0.90$ at $0.05$ intervals, and $n^*=0.93$. The first of these is well
within the dilute liquid region, and was used only to check that limit.
As explained in Ref.~({\no\VM}), it is necessary to include the parameter
$\lambda$ in order to represent the actual fluctuations through gaussian
noise. Its precise value is not crucial, since it essentially amounts
to a choice of the normalization of the static fluctuations $S(\vq)$,
but it clearly must be small since the density must always be positive.
We have taken here $\lambda=0.001$.
The remaining parameters $K$ and $\eta$ are functions of $n^*$ and may be
calculated as discussed in the previous Section.

We turn now to the very important question of data collection. We are
particularly concerned with ensuring that within statistical
error the averages collected be equilibrated, that is, stationary
in the time scales studied.
We find that, with the data collection procedure
as outlined below, the initial conditions that we use to begin the
integration of the equations of the motion are unimportant (except in that
they determine to some extent the duration of transient behavior) and we
usually take them to be a flat distribution of $n$ equal to its average
value, $n_0$, and vanishing currents.  We then monitor the current-current
correlations as a function of running time $t_0$. After a
relatively short time $t_{K}$ of order $10$ the current correlations
reach their equilibrium value as given by the equipartition theorem. It might
be tempting to assume that the density correlations have also equilibrated by
that time and such an assumption is sometimes made in MD work, but we find
that for our system at least this assumption does not hold.

To study the density correlations, we store, for running times $t_0 \ge
t_{K}$, where $t_0$ is the time measured from the initiation of the
computation, at a large number of periodic time bins, the products of the
form $\delta n(\vx,t_0) \delta n(\vx^\prime, t_0 + t)$
for all $\vx$ $\vx^\prime$. We then monitor
the spherically averaged spatial Fourier transform, $S(q,t,t_0)$ of
the  quantity:
$$ S(\vx,\vx^\prime,t,t_0) = <\delta n(\vx,t_0)\delta n(\vx^\prime,t_0+t)>
\eqnch$$ {\global\ES=\number\eqnumber}

\noindent where the average is understood to be over a number $n_b$ of
time bins separated by an interval $\Delta t$. The time range covered
by the averaging process is $t_R=n_b \Delta t$. In order for $S(q,t,t_0)$
to be an adequate approximation to the thermodynamic average $S(q,t)$ it
is required that it be not only independent of $t_0$, but also independent
of $t_R$ within statistical error. Dependence on the $t_0$ indicates the
presence of a transient. Dependence on $t_R$ indicates that the averaging
time is too short for ergodicity to hold. A very important point is that we
find that the minimum value of the transient time
for density fluctuations is not $t_K$, but it is of
the order of the slowest characteristic decay time $t^*$ in the system. As
discussed in the next Section, $t^*$ is a strongly increasing function of
density, and is much longer than the equilibration time for the kinetic energy.
Similarly, it is necessary
for the average to include a range $t_R$ of order of several times $t^*$.
We find that $S(q,t,t_0)$ at higher densities has considerable
oscillations over $t$ time ranges smaller than $t^*$.
As one increases the density an estimate for $t^*$ can be
found by extrapolation from the lower densities, since the behavior of $t^*$
with density turns out to obey the  Vogel-Fulcher law
$^{\no\VF}$.

Working within these constraints, obtaining statistically reliable results
still requires averaging over a large number of time bins which means very
large total running times. In addition, in order to eliminate any possibility
of spurious correlations due to a peculiar transient, we have repeated the
whole procedure three to five times at each density. The results presented
here correspond to a combined total of  between $1000$ and $3900$ bins,
depending on the density, at all densities  we present
results for except $n^*=0.75$ where a total
of $600$ was taken. These very large numbers, much larger than the
corresponding numbers in Ref.~(\no\VM), should be considered as
comparable to the `number of runs' in a standard simulation for a
nonequilibrium problem such as spinodal decomposition and lead to very
good quality data. The cost of obtaining the data rises
accordingly, of course: a total of 200 hours of Cray2 and Cray X/MP
time were required.

We have also verified that the quantity
$S(q,t=0)$ calculated following the above procedure and over the time ranges
just described is consistent
with the purely static result. To do this, first we
calculate the static result: we evaluate numerically, using
fast Fourier transforms, the discrete Fourier transform
$C(q)$ of $(\no\sectnf.\no\PYC)$ for a system of the size considered, and
we obtain from that result the static result $S_s(q)\propto
1/(1-n^*C(q))$
. To obtain this equation one must expand the first term in
$(\no\sectnf.\no\FB)$ to second order in $\delta n$. The comparison
between $S(q)$ and $S_s(q)$ is shown
in Fig. {\no\static}. We can see that the two results are in very good
agreement at the densities plotted.
The PY static values are known to be (see e.g. Fig 2 in Ref.({\no\AAY}))
in good agreement with MD results in this density range. Our results are
then also in agreement with MD in this limit.
At higher densities, the computed result represents a
higher degree of order than that calculated from the statics. This is due
to the fact that the expansion of the free energy just alluded to is not as
well justified at those densities. When averaged over time
scales of less than several times
$t^*$, $S(q,0)$ oscillates broadly about its stationary
value. Our results for $S(q,0)$ at all densities studied show that we
are dealing with a liquid-like state
here. If we use a commensurate value of
$\sigma$, or if we increase $n^*$ to $0.95$,
we find indications that crystallization begins. We plan to study this
crystallization question in future work.

The stability of our dynamically obtained results over long computational
runs, and their agreement with static results constitute a very stringent
check of the stability of our numerical algorithms.

To analyze our results, it is convenient to introduce the normalized quantity
$C(q,t)$ defined as:

$$ C(q,t) ={ {S(q,t)} \over{ S(q,t=0)} }
\eqnch $$ {\global\EC=\number\eqnumber}
where, we recall, we are dealing with angularly averaged quantities as
explained above with $q$ defined in $({\no\sectnr.\no\QE})$. We will then,
in the next Section,
characterize the decay for this quantity for all values of $q$.
By averaging $C(q,t)$  over a large effective number of runs
we are able to obtain results sufficiently smooth
to confidently
fit functional forms to the data as the next Section demonstrates.
\bigskip

\bigskip
\centerline{IV. RESULTS AND DISCUSSION }
\newsection {\global\sectna=\no\secno}

\bigskip

We now turn to the discussion of the normalized dynamic structure
factor  (see Eq.~$(\no\sectnr.\no\EC)$),
$C(q,t)$. We have first verified that at low densities ($n^*=0.5$), this
quantity decays exponentially at all wavevectors studied. We will consider
in what follows only the more interesting range $0.75\le n^* \le 0.93$.
We include in all of our fitting attempts all data up to either the
maximum time for which we have data at the particular values of $n^*$
and $q$ under consideration or up to the time where $C(q,t)$ is so small
that it fades into the noise. This occurs when $C(q,t)< 0.025$ except
in some very few cases where the data becomes noisy at the $0.05$ level.
As in Ref.~({\no\VM}) the statistical noise seems to have an additive
component which causes the relative errors to increase when $C(q,t)$
becomes smaller. We recall that the relation between our time units and
Enskog collision time is given by $({\no\sectnf}.\no\TU)$. The factor relating
the two inverse times varies from $1.52$ at $n^*=0.75$ to $1.90$ at
$n^*=0.93$. The relation between our unit of length and $\sigma$ is the
trivial factor of $\sigma/h=4.6$.

We begin by attempting a fit to a stretched exponential form:
$$C(q,t)=e^{-(t/\tau_0)^\beta}   \eqnch $$ {\global\ST=\no\eqnumber}
\noindent where the parameters $\tau_0$ and $\beta$ are functions of $n^*$
and $q$. This step is motivated in part by the expectation from preliminary
inspection of the data that there is a wide region of $q$ and $n^*$ values
for which this form is adequate. Indeed we find that
for a considerable part of
the data particularly in the lower density region this turns out to be a
satisfactory fit. Even in the cases where the data is not
well fitted by the form of Eq.~$(\no\sectna.\no\ST)$, the number
$\tau_0(q,n^*)$
still gives a useful figure of merit or overall estimate of the decay time.
The results of this fit are in Tables I and II. We have indicated in these
Tables the cases where the fit to the above form is a good  fit to the data
and those in which it is actually not the best fit by enclosing the latter
cases in parentheses. The quality of the fits can be judged visually
and by the $\chi^2$ values that we obtain.

The salient points of the results in Table I are apparent: the characteristic
time is at constant density a very strong function of $q$, and it has a
maximum at the wavevector shells close to where the static structure factor
$S(q)$ has its maximum. A less well defined secondary maximum at wavevectors
close to the second maximum of $S(q)$ can also be
discerned.  The characteristic time is not however of the simple form
$\tau_0(q,n^*)\propto \eta S(q)$ as one might guess naively: it is neither
proportional to $S(q)$ at constant density (recall $\eta$ depends on the
density only) nor proportional to $\eta$ at constant $q$. This indicates
as might be expected,
a q-dependent renormalization of the transport characteristic times.
This is more marked at higher densities.
The stretching parameter $\beta$ (Table II)
remains relatively close to unity, although
stretching is evident as one increases the density particularly at shorter
wavelengths. We note that the entries in the Table corresponding to the smaller
values of $\beta$ are in most cases  not good fits to the data, as we
shall discuss below.

It is apparent from our tabulated results that the slowest overall decay
is found, in our spherically averaged results, at the shell number $Q=8$,
near the main peak in $S(q)$. We consider then the main decay time $\tau_0$ at
that wavevector, as representing the slowest transport time at that
density $t^*$. We plot in Fig.~({\no\FVF}) this quantity as extracted
from Table I  as a function
of density. We have fitted this quantity to the Vogel-Fulcher $^{\no\VF}$
law $({\no\sectni}.\no\VFE)$ which when the density, rather than the
temperature, is the externally controlled variable, takes the form:
$$ t^*(n^*)=\alpha e^{\gamma/(v-v_c)}  \eqnch$$
where $v \equiv 1/n^*$. As shown in the Figure, we find an excellent fit
with a value of $v_c$ corresponding to $n_c^*=1.23$. This is in excellent
agreement with the  analysis of MD data of Ref.$({\no\WA})$ where
the result $n_c^*=1.21$ was obtained. A three parameter fit to a power
law is also adequate, but the quality of the fit is not quite as good.

We now turn to the decay modes for which the stretched exponential is not
a satisfactory fit.
We have examined the data very carefully and attempted a large number of fits
to different proposed decay forms. We find that the reason for the failure of
$(\no\sectna.\no\ST)$ is usually
that the decay is a two step process, or in some cases that the data
appear to decay to a finite constant in the time range studied. Accordingly, we
have found that the best form that fits  this portion of the data is :
$$C(q,t)=(1-f) e^{-(t/\tau)^\beta} + f e^{-t/\tp} \eqnch $${\global\LU
=\no\eqnumber}
where it is understood that the time scales are well separated, $\tp
\gg \tau$. One can also include a stretching exponent in the second decay,
but we have found it usually unnecessary
and we were leery of multiplying the number of fitting parameters.
We note that the form $(\no\sectna.\no\ST)$ is a particular case of
$(\no\sectna.\no\LU)$ with $f=0$
or, alternatively, $\tau\approx\tp$. It also represents a decay to a
constant (quasinonergodic behavior) whenever $\tp=\infty$, which of
course should be understood as meaning
that the second decay time is beyond the region
fitted. Finally, with $f$ finite, $(\no\sectna.\no\LU)$ yields a region of
apparent power law decay of the von Schweidler form
when $\tau \ll t \ll \tp $ and the time
scales are widely separated. We find that this very general form fits well
(again, as measured by the appropriate reduced $\chi^2$ values and visually)
all cases where as indicated by the parentheses in Tables I and II
the single decay fit was not satisfactory. The
parameters $\tau$, $\tp$, and $f$ are given in Table III. The parameter
$\beta$ turned out to be very close to unity in all of these cases. In Figures
($\no\ddd$) and ($\no\ddc$) we show a wide selection of our results and the
corresponding best fits, according to the parameters in Tables I-III.
For every density we have two Figures, which display the decay of
$C(q,t)$ at that density for a sampling of wavevector values.
The values shown are $Q=6,8,10,12,13,14$.
These include all $Q$ values where the decay is not of the form
$({\no\sectna}.\no\ST)$. Values skipped have been omitted to avoid
excessive clutter. Keeping in mind
the Figures and the three Tables,
we see very clear trends. First, as one increases
the density, nonexponential behavior crops out first at relatively larger
wavevectors, that is, shorter distances, which is physically sensible and
was observed also in Ref.~(\no\VM).
Specifically, this behavior appears first at values beyond the main peak in
the structure factor, in the region of the second peak. As the density
increases further the beginning of the same phenomenon is observed
at smaller wavevectors at or near the main peak in $S(q)$. This
phenomenon appears to be related to the $q$ dependence of $\beta$ found in
mode coupling calculations\ref{M. Fuchs, I. Hofacker, and A. Latz,
Phys. Rev. {\bf A45}, 898, (1992).} . The
characteristic times increase strongly with density, and so does their
relative separation. They are also strongly $q$ dependent. On the other
hand, $f$ appears to depend rather weakly on $n^*$ but more strongly on
$q$, and its $q$ dependence appears also to correlate with that of
the statics. From the point of view of $(\no\sectna.\no\LU)$ it appears that
the lower density stretched exponential form
$(\no\sectna.\no\ST)$ should be viewed as arising
from a confluence of the time scales, rather than from the vanishing of $f$.

We monitor also the value of the mean field free energy during the computation.
We find that the system fluctuates around a liquid like state minimum.
Thus, our simulations establish that the first three or four orders of
magnitude in the growth of the relaxation time in the hard sphere system
arise from the effects of nonlinearities associated with the realistic free
energy used.

It is very instructive to discuss our results in connection with MD
work, with mode coupling theoretical work on glassy systems, and
with experiment. Some caution is
required, however, since it must be kept in mind that we are dealing here with
relatively low densities: the degree of `supercooling' that our choices of $N$
and $\sigma$ achieve is limited in comparison with what can be done
experimentally. We have not used other procedures, such as mixing spheres of
two different sizes, which might allow further increases
in the density. There are a variety of decay regimes and forms reported
in the literature for glassy systems. A scenario in favor of which some
consensus has gathered $^{\no\KM}$
is described in the Introduction.
We believe that our results at higher densities and shorter
distances, in the region where the full form $(\no\sectna.\no\LU)$ is obeyed
might be interpreted within the above scheme as representing a first
decay which combines phonon and $\beta$ -relaxation, leading eventually to the
second decay which might then be identified with $\alpha$ -relaxation. In
some cases only the quasinonergodic regime is reached in our computations. The
von Schweidler regime can arise, as explained above, as a precursor of the
second decay. The phonon assisted decay does not appear as a separate regime
in our data, which is possibly due to the coarse graining nature of Langevin
dynamics with respect to microscopic times. We have verified that an inverse
power law does not fit the earlier portion of our data as well as a stretched
exponential form.

We have already mentioned the agreement of our results for $t^*$ and
$S(q)$ with the MD results of Refs. ({\no\WA}) and ({\no\AAY})
respectively.  Comparison with MD work on glass dynamics
is complicated by the fact that such
MD results are for different systems, such as binary soft sphere
mixtures (see e.g. Refs. ({\no\UY}) and
({\no\BRH})) or Coulombic systems ({\no\SIG}) . Obviously, detailed
comparisons will have to await further work. However,
as results for different MD systems are qualitatively consistent with
each other, a qualitative comparison with our results is possible. Although
comparison of MD and Langevin time scales is in general a vexing
problem, for our purposes here it might be adequate to simply assign to
our phonon-based time unit a value of order $0.1$ ps. We then find that
the results of Ref.({\no\UY}) correspond to a shorter time range than our
higher
density results, while those of Refs.({\no\BRH}) and ({\no\SIG}) are
comparable.
Neither Ref.({\no\BRH}) which covers higher densities than ours, nor
Ref.({\no\SIG}) follow $C(q,t)$ until it decays to zero or to a small value, as
we do, and they assume that their systems have reached equilibrium when
the kinetic energy equilibrates,
an assumption which we checked and we found, for our
dynamics, erroneus. We do not know if this is also the case for
MD. Despite all of these caveats, our results
are consistent with those of both Refs. ({\no\BRH}) and ({\no
\SIG}).
They observe first a glitch at very early times, which they attribute to
phonons, and then two relaxation modes separated by a slow region, which
they identify with $\beta$ and $\alpha$- relaxations. They do not
attempt to quantitatively characterize the $\beta$ regime by either a power
law or an exponential, but they do obtain an exponential form
(stretched at higher densities) for the second decay. As stated above,
our data does not show any early time glitch, which could be due either to
Langevin coarse graining, or to the glitch being an artifact of poor system
equilibration. Except for this, our results are consistent with MD
in that a naive extrapolation of our results to higher densities would
predict that the time scales $\tau$ and $\tp$ in $(\no\sectna.\no\LU)$
become more separated, and possibly larger values of $f$ appear. We have
verified that the MD data of Ref.({\no\SIG}) can in fact be fit to the form
$(\no\sectna.\no\LU)$ with appropriate parameter values. Thus we conclude
that our results are consistent with MC as well as MD.

\bigskip
%\vfill\eject
\centerline{ACKNOWLEDGMENTS}
\bigskip
We are indebted to G.F. Mazenko for many and very useful discussions about
this problem. L.M.L. was supported in part by a Department of Education
Fellowship. C.D. thanks the School of Physics at the University of Minnesota
and the Minnesota Supercomputer Institute for support and hospitality
and S. Ramaswamy for useful discussions. This
work was supported in part by  grants from the Cray Research
Corporation and the Minnesota Supercomputer Institute.
\vfill
\eject
\noindent\reference
\vfill\eject
\noindent{\bf FIGURE CAPTIONS}

\parindent0pt

{\bf Fig.~{\no\static}.} The static $S_s(Q)$ calculated as  explained in the
text, compared with the numerical
results for $S(Q)=S(Q,0)$. Both quantities are spherical averages
plotted vs. wavevector as given by the shell number $Q$ defined in the
text. The curves shown are for $n^*=0.75$ ($S_s(Q)$ solid line, $S(Q,0)$
medium dashes) and for $n^*=0.80$ (static results short dashes, dynamic
results shown as dots).

{\bf Fig.~{\no\FVF}.} The characteristic decay time, $t^*$ plotted as a
function of density $n^*$. Data is fitted to a Volger-Fulcher form as
explained in the text.

{\bf Fig.~{\no\ddd}.} The normalized correlation function $C(q,t)$
plotted vs. t for $Q$= 14 (top curve),12, and 6 (bottom curve).
The data and the best functional fits (smooth curves) as explained
in the text are shown for (a) $n^*=0.75$ , (b) $n^*=0.80$ ,
(c) $n^*=0.85$, (d) $n^*=0.90$ , and (e) $n^*=0.93$ .

{\bf Fig.~{\no\ddc}.} As in the previous figure but for
$Q$= 8 (top curve) ,13, and 10 (bottom curve) at $t=100$.

\vfill
\eject

\centerline{TABLE CAPTIONS}

{\bf Table I}.- The parameter  $\tau_0$ of Eq.(4.1) as a function of density
as given by $n^*$ and wavevector as given by shell number $Q$. The parentheses
indicate data sets for which the fit is not satisfactory. These cases
are fitted by Eq. (4.3) and Table III.

{\bf Table II}.- The parameter  $\beta$  of Eq.(4.1) as a function of density
as given by $n^*$ and wavevector as given by shell number $Q$.

{\bf Table III}.- The parameters $\tau$,$\tp$, and $f$ of Equation (4.3),
for the appropriate range of $n^*$ and $Q$ values, (see text and Tables
I and II).  The values of $\tp$ indicated by infinity must be interpreted
as being beyond the time window fitted. The three daggered values of
$\tp$ indicate cases where an exponent $\beta^\prime$ different from
unity was required. In these cases $\beta$ was set to unity, so the number of
parameters was the same, but in the other cases the unrestricted $\beta$
remained close to unity.

\vfill
\eject

\centerline{Table I}

\table{$$\vbox{\offinterlineskip
\hrule
\halign{&\vrule#&
 \strut\quad\hfil#\quad\cr
height2pt&\omit&&\omit&&\omit&&\omit&&\omit&&\omit&\cr
&$Q\backslash
n^*$\hfil&&0.75\hfil&&0.80\hfil&&0.85\hfil&&0.90\hfil&&0.93\hfil&\cr
height2pt&\omit&&\omit&&\omit&&\omit&&\omit&&\omit&\cr
\noalign{\hrule}
height2pt&\omit&&\omit&&\omit&&\omit&&\omit&&\omit&\cr
height2pt&\omit&&\omit&&\omit&&\omit&&\omit&&\omit&\cr
&  6&&   32 &&   37 &&   38 &&   39 &&   40 &\cr
height2pt&\omit&&\omit&&\omit&&\omit&&\omit&&\omit&\cr
height2pt&\omit&&\omit&&\omit&&\omit&&\omit&&\omit&\cr
&  7&&   81 &&  102 &&  133 &&  159 &&  156 &\cr
height2pt&\omit&&\omit&&\omit&&\omit&&\omit&&\omit&\cr
height2pt&\omit&&\omit&&\omit&&\omit&&\omit&&\omit&\cr
&  8&&  130 &&  183 &&  424 &&(1140)&&(2271)&\cr
height2pt&\omit&&\omit&&\omit&&\omit&&\omit&&\omit&\cr
height2pt&\omit&&\omit&&\omit&&\omit&&\omit&&\omit&\cr
&  9&&  103 &&  131 &&  208 &&  344 &&  501 &\cr
height2pt&\omit&&\omit&&\omit&&\omit&&\omit&&\omit&\cr
height2pt&\omit&&\omit&&\omit&&\omit&&\omit&&\omit&\cr
& 10&&   71 &&   87 &&  112 &&  149 &&  192 &\cr
height2pt&\omit&&\omit&&\omit&&\omit&&\omit&&\omit&\cr
height2pt&\omit&&\omit&&\omit&&\omit&&\omit&&\omit&\cr
& 11&&   93 &&  108 &&  125 &&  173 &&  186 &\cr
height2pt&\omit&&\omit&&\omit&&\omit&&\omit&&\omit&\cr
height2pt&\omit&&\omit&&\omit&&\omit&&\omit&&\omit&\cr
& 12&&  68 && (105)&& (118)&& (157)&& (186)&\cr
height2pt&\omit&&\omit&&\omit&&\omit&&\omit&&\omit&\cr
height2pt&\omit&&\omit&&\omit&&\omit&&\omit&&\omit&\cr
& 13&& (128)&& (197)&& (245)&& (347)&& (430)&\cr
height2pt&\omit&&\omit&&\omit&&\omit&&\omit&&\omit&\cr
height2pt&\omit&&\omit&&\omit&&\omit&&\omit&&\omit&\cr
& 14&& (134)&& (150)&& (247)&& (221)&& (301)&\cr
height2pt&\omit&&\omit&&\omit&&\omit&&\omit&&\omit&\cr
height2pt&\omit&&\omit&&\omit&&\omit&&\omit&&\omit&\cr}
\hrule}$$}

\vfill
\eject

\centerline{Table II}

\table{$$\vbox{\offinterlineskip
\hrule
\halign{&\vrule#&
 \strut\quad\hfil#\quad\cr
height2pt&\omit&&\omit&&\omit&&\omit&&\omit&&\omit&\cr
&$Q\backslash
n^*$\hfil&&0.75\hfil&&0.80\hfil&&0.85\hfil&&0.90\hfil&&0.93\hfil&\cr
height2pt&\omit&&\omit&&\omit&&\omit&&\omit&&\omit&\cr
\noalign{\hrule}
height2pt&\omit&&\omit&&\omit&&\omit&&\omit&&\omit&\cr
height2pt&\omit&&\omit&&\omit&&\omit&&\omit&&\omit&\cr
&  6&&  0.87 &&    0.84 &&    0.82 &&    0.89 &&    0.84 &\cr
height2pt&\omit&&\omit&&\omit&&\omit&&\omit&&\omit&\cr
height2pt&\omit&&\omit&&\omit&&\omit&&\omit&&\omit&\cr
&  7&&  0.95 &&    1.00 &&    0.82 &&    0.93 &&    0.90 &\cr
height2pt&\omit&&\omit&&\omit&&\omit&&\omit&&\omit&\cr
height2pt&\omit&&\omit&&\omit&&\omit&&\omit&&\omit&\cr
&  8&&  0.98 &&    0.99 &&    0.84 &&    (0.78)&&    (0.94)&\cr
height2pt&\omit&&\omit&&\omit&&\omit&&\omit&&\omit&\cr
height2pt&\omit&&\omit&&\omit&&\omit&&\omit&&\omit&\cr
&  9&&  0.96 &&    1.00 &&    1.01 &&    0.94 &&   0.89  &\cr
height2pt&\omit&&\omit&&\omit&&\omit&&\omit&&\omit&\cr
height2pt&\omit&&\omit&&\omit&&\omit&&\omit&&\omit&\cr
& 10&&  1.00 &&    0.95 &&    1.04 &&    1.03 &&    0.99 &\cr
height2pt&\omit&&\omit&&\omit&&\omit&&\omit&&\omit&\cr
height2pt&\omit&&\omit&&\omit&&\omit&&\omit&&\omit&\cr
& 11&&  0.84 &&    0.91 &&    0.94 &&    0.90 &&    0.93 &\cr
height2pt&\omit&&\omit&&\omit&&\omit&&\omit&&\omit&\cr
height2pt&\omit&&\omit&&\omit&&\omit&&\omit&&\omit&\cr
& 12&& 0.83 &&    (0.66)&&    (0.78)&&    (0.72)&&    (0.56)&\cr
height2pt&\omit&&\omit&&\omit&&\omit&&\omit&&\omit&\cr
height2pt&\omit&&\omit&&\omit&&\omit&&\omit&&\omit&\cr
& 13&&  (0.74)&&    (0.69)&&    (0.67)&&    (0.66)&&    (0.60)&\cr
height2pt&\omit&&\omit&&\omit&&\omit&&\omit&&\omit&\cr
height2pt&\omit&&\omit&&\omit&&\omit&&\omit&&\omit&\cr
& 14&&  (0.69)&&    (0.77)&&    (0.47)&&    (0.54)&&    (0.32)&\cr
height2pt&\omit&&\omit&&\omit&&\omit&&\omit&&\omit&\cr
height2pt&\omit&&\omit&&\omit&&\omit&&\omit&&\omit&\cr}
\hrule}$$}
\bigskip

\vfill
\eject

\centerline{Table III}

\table{$$\vbox{\offinterlineskip
\hrule
\halign{&\vrule#&
 \strut\quad\hfil#\quad\cr
height2pt&\omit&&\omit&&\omit&&\omit&&\omit&&\omit&&\omit&\cr
&$Q\backslash n^*$\hfil&&
para.\hfil&&0.75\hfil&&0.80\hfil&&0.85\hfil&&0.90\hfil&&0.93\hfil&\cr
height2pt&\omit&&\omit&&\omit&&\omit&&\omit&&\omit&&\omit&\cr
\noalign{\hrule}
height2pt&\omit&&\omit&&\omit&&\omit&&\omit&&\omit&&\omit&\cr
height2pt&\omit&&\omit&&\omit&&\omit&&\omit&&\omit&&\omit&\cr
&  8&& $\tau$ &&   -  &&   -  &&   -  &&  561 &&  942 &\cr
height2pt&\omit&&\omit&&\omit&&\omit&&\omit&&\omit&&\omit&\cr
height2pt&\omit&&\omit&&\omit&&\omit&&\omit&&\omit&&\omit&\cr
&   && $\tau^\prime$&&   -  &&   -  &&   -  && 3115$\dag$&& 4231$\dag$&\cr
height2pt&\omit&&\omit&&\omit&&\omit&&\omit&&\omit&&\omit&\cr
height2pt&\omit&&\omit&&\omit&&\omit&&\omit&&\omit&&\omit&\cr
&   && $f$ &&   -  &&   -  &&   -  && 0.34 && 0.45 &\cr
height2pt&\omit&&\omit&&\omit&&\omit&&\omit&&\omit&&\omit&\cr
\noalign{\hrule}
height2pt&\omit&&\omit&&\omit&&\omit&&\omit&&\omit&&\omit&\cr
height2pt&\omit&&\omit&&\omit&&\omit&&\omit&&\omit&&\omit&\cr
& 12&& $\tau$ &&   -  &&   83 &&   98 &&  131 &&  149 &\cr
height2pt&\omit&&\omit&&\omit&&\omit&&\omit&&\omit&&\omit&\cr
height2pt&\omit&&\omit&&\omit&&\omit&&\omit&&\omit&&\omit&\cr
&   && $\tau^\prime$&&   -    &&   $\infty$  &&   $\infty$  &&   $\infty$  &&
$\infty$  &\cr
height2pt&\omit&&\omit&&\omit&&\omit&&\omit&&\omit&&\omit&\cr
height2pt&\omit&&\omit&&\omit&&\omit&&\omit&&\omit&&\omit&\cr
&   && $f$ &&   -  && 0.08 && 0.07 && 0.07 && 0.09 &\cr
height2pt&\omit&&\omit&&\omit&&\omit&&\omit&&\omit&&\omit&\cr
\noalign{\hrule}
height2pt&\omit&&\omit&&\omit&&\omit&&\omit&&\omit&&\omit&\cr
height2pt&\omit&&\omit&&\omit&&\omit&&\omit&&\omit&&\omit&\cr
& 13&& $\tau$ &&   51 &&   64 &&  104 &&  107 &&  171 &\cr
height2pt&\omit&&\omit&&\omit&&\omit&&\omit&&\omit&&\omit&\cr
height2pt&\omit&&\omit&&\omit&&\omit&&\omit&&\omit&&\omit&\cr
&   && $\tau^\prime$&&  239 &&  397 &&  615 &&  761 && 1210 &\cr
height2pt&\omit&&\omit&&\omit&&\omit&&\omit&&\omit&&\omit&\cr
height2pt&\omit&&\omit&&\omit&&\omit&&\omit&&\omit&&\omit&\cr
&   && $f$ && 0.55 && 0.56 && 0.44 && 0.53 && 0.42 &\cr
height2pt&\omit&&\omit&&\omit&&\omit&&\omit&&\omit&&\omit&\cr
\noalign{\hrule}
height2pt&\omit&&\omit&&\omit&&\omit&&\omit&&\omit&&\omit&\cr
height2pt&\omit&&\omit&&\omit&&\omit&&\omit&&\omit&&\omit&\cr
& 14&& $\tau$ &&    94 &&  112 &&  135 &&  159 &&  219 &\cr
height2pt&\omit&&\omit&&\omit&&\omit&&\omit&&\omit&&\omit&\cr
height2pt&\omit&&\omit&&\omit&&\omit&&\omit&&\omit&&\omit&\cr
&   && $\tau^\prime$&&  702$\dag$&&   $\infty$  &&   $\infty$  &&   $\infty$
&&  $\infty$  &\cr
height2pt&\omit&&\omit&&\omit&&\omit&&\omit&&\omit&&\omit&\cr
height2pt&\omit&&\omit&&\omit&&\omit&&\omit&&\omit&&\omit&\cr
&   && $f$ &&  0.13 && 0.14 && 0.18 && 0.16 && 0.14 &\cr
height2pt&\omit&&\omit&&\omit&&\omit&&\omit&&\omit&&\omit&\cr
height2pt&\omit&&\omit&&\omit&&\omit&&\omit&&\omit&&\omit&\cr}
\hrule}$$}

\bye